# Preemptive Spatiotemporal Trajectory Adjustment for Heterogeneous Vehicles in Highway Merging Zones

YuanLi[a,1]，Xiaoxue Xu[a,1]， XiangDong[a,1]，Junfeng Hao[b], TaoLi[a], Sana Ullah[a], Chuangrui Huang[a]，Junjie Niu[a], Ziyan Zhao[a], Ting Peng[a]*

[a]Key Laboratory for Special Area Highway Engineering of the Ministry of Education, Chang'an University, Xi'an, China

[b] China Railway No.7 Bureau Group No.3 Engineering Co., Ltd. Xi'an, China

*Corresponding Author(s): E-Mail: t.peng@ieee.org

Contributing authors: liyuan_mm@chd.edu.cn; 2112534668@qq.com;1290740787@qq.com; 935524208@qq.com; lt1186166770@163.com; 3493463476@qq.com;1304584578@qq.com; 3312487631@qq.com;253841200@qq.com

**Abstract:** Aiming at the problem of driver's perception lag and low utilization efficiency of space-time resources in expressway ramp confluence area, based on the preemptive spatiotemporal trajectory Adjustmentsystem, from the perspective of coordinating sptio-time resources, the reasonable value of safe space-time distance in trajectory pre-preparation is quantitatively analyzed. The minimum safety gap required for ramp vehicles to merge into the mainline is analyzed by introducing double positioning error and spatio-time trajectory tracking error. A merging control strategy for autonomous driving heterogeneous vehicles is proposed, which integrates vehicle type, driving intention and safety spatio-time distance. The specific confluence strategies of ramp target vehicles and mainline cooperative vehicles under different vehicle types are systematically expounded. A variety of traffic flow and speed scenarios are used for full combination simulation. By comparing the time-position-speed diagram, the vehicle operation characteristics and the dynamic difference of confluence are qualitatively analyzed, and the average speed and average delay are used as the evaluation indexes to quantitatively evaluate the performance advantages of the preemptive cooperative confluence control strategy. The results show that the maximum average delay improvement rates of mainline and ramp vehicles are 90.24 % and 74.24 %, respectively. The proposed strategy can effectively avoid potential vehicle conflicts and emergency braking behaviors, improve driving safety in the confluence area, and show significant advantages in driving stability and overall traffic efficiency optimization.

**Keywords**：Connected autonomous driving environment, vehicle heterogeneity, on-ramp merging, preemptive collaborative control

## 1 Introduction

On-ramp merging areas serve as bottlenecks on highways, featuring more complex traffic flow characteristics than other sections, which renders them prone to traffic congestion and accidents. According to statistics, accidents in highway on-ramp areas account for more than 30% of total highway accidents in China each year (Qi, Liu et al. 2022). Especially under high traffic volumes, drivers cannot accurately judge merging opportunities, thereby triggering significant safety hazards. Therefore, optimizing on-ramp merging strategies is crucial for enhancing overall traffic efficiency and safety.

Scholars at home and abroad have conducted extensive research on highway on-ramp merging problems based on autonomous driving and vehicle networking technologies, but the research mainly focuses on homogeneous traffic flow environments. Even in scenarios involving heterogeneous traffic flows with mixed Connected and Automated Vehicles (CAVs) and Human-Driven Vehicles (HDVs) (el abidine Kherroubi, Aknine et al. 2021;Hao, Gong et al. 2023;Liu, Zhuang et al. 2023;Wei, Xinyue et al. 2023), it is generally assumed that vehicle types are consistent, with less in-depth exploration of the impact of vehicle heterogeneity on the merging process. With the acceleration of intelligent transformation of passenger and freight vehicles, and the evolution of traffic patterns toward diversified mixed traffic, the heterogeneity of vehicle types at highway on-ramp entrances has become increasingly prominent. This assumption not only contradicts the current actual traffic composition but also fails to adapt to the trend of vehicle intelligent transformation. The differences in dynamic characteristics and vehicle sizes between medium/large and small vehicles (Liu, Zhao et al. 2023) not only exacerbate interference between vehicles but may also lead to reduced traffic capacity, decreased traffic efficiency, and increased risks of traffic congestion and accidents. Statistical data show that when medium/large vehicles are involved in accidents, the proportion of severe accidents accounts for 74.30% of total accidents, while this proportion is only 48.03% in accidents not involving medium/large vehicles (Wu 2020). In addition, most existing studies use single-mainline single-ramp scenarios (Fang, Min et al. 2022;Jing, Hui et al. 2022;Shi, Li et al. 2023), and research on cooperative merging control of vehicles in multi-lane mainline environments is relatively scarce, limiting the applicability and promotion value of relevant research findings in more complex traffic environments.

To address the challenge that cooperative control in current traffic systems is constrained by local information games, Li, Li et al. (2024) proposed a Preemptive Holistic Collaborative System (PHCS), which constructs a proactive collaborative framework based on the intentions of embodied agents. By real-time perceiving the environmental state, the states of each embodied agent, and their future action goals, the system pre-plans spatiotemporal actions and forms a consensus, thereby achieving efficient and safe collaboration in complex systems. It breaks through the limitation of traditional embodied agents relying on local feedback for collaboration, and provides an innovative solution to the collaboration problem among system participants through a closed-loop mechanism of "intention-driven - global optimization - conflict pre-resolution". When applied to road traffic, the system can pre-plan spatiotemporal trajectories based on the intentions of all vehicles, eliminating conflicts between vehicles from the root and completely changing the traditional traffic management mode. Under high traffic volumes, the system can reduce the average delay time of vehicles by up to 97.96%, decrease fuel consumption by approximately 6.01%, increase road capacity by more than 4 times, and eliminate more than 90% of traffic accidents. However, this study has limitations: its research scenario is also limited to single-mainline single-ramp, and does not consider the optimization of merging strategies in vehicle heterogeneity and multi-mainline environments.

In summary, how to optimize multi-mainline on-ramp merging strategies in the connected autonomous driving environment, especially to effectively address the collaborative complexity brought by mixed vehicle types, and to achieve efficient allocation of spatiotemporal resources in multi-mainline environments, has become a key problem to be solved in the field of intelligent transportation.

To address the challenges posed by mixed vehicle types and fully leverage the application potential of intelligent connected autonomous driving technologies and real-time information sharing in the on-ramp merging process, it is urgent to explore merging strategies for efficient allocation of spatiotemporal resources and construct a collaborative control framework that balances spatiotemporal safe spacing and adapts to mixed traffic environments with different vehicle types. Through this collaborative control framework, the spatiotemporal allocation of traffic flow can be optimized, the operational efficiency of the traffic system can be improved, and the safety in the on-ramp merging process can be effectively enhanced, reducing potential traffic conflicts and accident risks.

Considering spatiotemporal errors, this paper focuses on the merging scenario of double-mainline single-ramp on highways, and relies on the information sharing mechanism in the preemptive spatiotemporal trajectory system to achieve efficient allocation of spatiotemporal resources under multi-mainline conditions, avoiding resource waste caused by local information games. It contributes a preemptive collaborative merging control strategy for heterogeneous traffic flows (pre-planning spatiotemporal trajectories) to optimize merging efficiency and traffic safety. To the best of our knowledge, no existing study has comprehensively addressed these issues within a systematic and rigorous methodological framework, which distinguishes it from all other works in the literature.

First, the research on preemptive collaborative merging control strategy is carried out. Based on the quantitative analysis of spatiotemporal trajectory tracking errors and positioning errors, the reasonable value of safe spatiotemporal spacing in pre-planned trajectories is analyzed, and the minimum safe gap required for on-ramp vehicles to merge into the mainline is further clarified. The calculation method of safe spatiotemporal spacing is the sum of twice the positioning error and the spatiotemporal trajectory tracking error. On this basis, the minimum safe gap required for on-ramp vehicles to merge into the mainline is further clarified, which is calculated as the sum of twice the safe spatiotemporal spacing and the body length of on-ramp vehicles.

In addition, aiming at the merging problem of heterogeneous vehicles in double-mainline single-ramp on highways, to minimize the risk of traffic conflicts and optimize the utilization of spatiotemporal resources, a preemptive collaborative merging strategy integrating vehicle types, driving intentions, and spatiotemporal safe spacing is proposed.

Finally, the simulation experiment research on the merging scenario of double-mainline single-ramp heterogeneous traffic flow is carried out. Based on the joint simulation of SUMO and Python, the rationality and effectiveness of the proposed preemptive collaborative merging strategy combining vehicle types, driving intentions, and spatiotemporal safe spacing are verified. By constructing a simulation environment for the double-mainline single-ramp merging area of highways and designing test schemes, the performance differences between the strategy and the uncontrolled case are compared and analyzed from three aspects: average delay time, average driving speed, and total fuel consumption, to evaluate the performance advantages and applicability of the strategy.

The structure of this paper is as follows. Section 2 briefly reviews the previous literature and identifies research gaps. Section 3 carries out the research on preemptive collaborative merging control strategy based on the preemptive collaborative merging system architecture. Aiming at the merging problem of heterogeneous vehicles in double-mainline single-ramp on highways, a preemptive collaborative merging control strategy integrating vehicle types, driving intentions, and spatiotemporal safe spacing is proposed to reduce the risk of traffic conflicts and optimize the utilization of spatiotemporal resources, and the specific merging strategies for on-ramp target vehicles and mainline cooperative vehicles under different vehicle type combinations are systematically expounded. Section 4 evaluates the effectiveness of the performance advantages of the preemptive collaborative merging control strategy through numerical experiments. Finally, Section 5 summarizes the entire research work and main findings.

## 2  Literature review

With the rapid development of intelligent and connected technologies, vehicles are evolving from assisted driving to autonomous driving, and shifting from single-vehicle intelligence to multi-vehicle collaboration. However, current autonomous driving technologies still struggle to accurately and real-time acquire the states and driving intentions of all surrounding vehicles in practical applications, thus limiting the further improvement of cooperative merging control effects.

Research on Intelligent Connected Autonomous Driving Technologies. Connected and Automated Vehicles (CAVs), by integrating automation and connectivity technologies, have been widely applied to improve road traffic efficiency, enhance safety and stability, and reduce energy consumption and emissions. The integrated development direction of intelligence and connectivity has been widely recognized. Strategic documents such as the U.S. Intelligent Transportation System Strategic Plan 2020–2025 (JPO), the EU Connected, Cooperative and Automated Mobility Roadmap (ERTRAC), China's Smart Vehicle Innovation Development Strategy(Commission 2020), and Intelligent Connected Vehicle Technology Roadmap 2.0 (Guo 2020)provide forward-looking guidance for enhancing intelligent driving capabilities through networked collaboration. Among them, Intelligent Connected Vehicle Technology Roadmap 2.0 formulates phased development goals for passenger vehicles and freight vehicles, revealing the possibility of collaborative mixed operation between autonomous passenger vehicles and autonomous freight vehicles in highway environments. Against the backdrop of countries promoting CAV development, many technology companies and automakers have invested in autonomous driving R&D, accelerating the intelligent and connected progress of vehicles through cross-industry cooperation (Wang 2022). Meanwhile, in highway transportation scenarios, autonomous freight vehicles have lower perception and prediction requirements than autonomous passenger vehicles (Wang, Gao et al. 2022), with fixed driving routes and relatively stable speeds, making them easier to implement. In recent years, domestic and foreign research on autonomous freight vehicles in highway scenarios has mainly focused on the optimization and application of platooning technologies. (De Curtò, de Zarzà et al. 2023) proposed a drone-based Decentralized Truck Platooning (DDTP) method, integrating Model Predictive Control (MPC) and Unscented Kalman Filter (UKF) technologies to enhance the efficiency and safety of highway platooning. Chen, Zhou et al. (2023)developed a Platoon-MAPPO algorithm based on multi-agent reinforcement learning for truck platoon control in highway on-ramp scenarios, improving overall platoon efficiency while ensuring traffic safety, reducing average energy consumption by 14.8% and road occupancy by 43.3%. To make platoon strategies more compatible with actual road traffic characteristics, Zhao,Pang (2024) proposed independent and cooperative platooning strategies for intelligent connected vehicles based on platoon intensity and penetration rate, aiming to improve road capacity by addressing the mixed traffic characteristics of intelligent connected trucks, cars, and human-driven trucks/cars.

Research on On-Ramp Merging Cooperation Strategies. Highway on-ramp merging areas are typical bottlenecks in road traffic. CAVs, through real-time communication and precise trajectory control technologies, are expected to significantly reduce or avoid traffic conflicts in on-ramp merging areas. Current research on cooperative merging strategies for CAVs at on-ramps mainly adopts three approaches: rule-based, optimization-based, and learning-based methods.

Rule-based methods rely on predefined rules. Ding, Li et al. (2020) proposed a rule-based adjustment algorithm to obtain near-optimal merging sequences with low computational cost, introducing virtual vehicle mapping and energy efficiency methods to address acceleration rate issues. Yang, Dong et al. (2023) developed a ramp cooperative control strategy for CAVs based on virtual platooning. Wang, Gong et al. (2024)proposed a Distributed Learning-based Iterative Optimization (DLIO) method to optimize merging trajectories, introducing a Heuristic Monte Carlo Tree Search (HMCTS) algorithm based on improved search principles to solve the minimum time passage sequence problem.

Optimization-based methods adopt a global perspective, using mathematical modeling or optimization algorithms with objective functions and constraints. Examples include mixed-integer nonlinear programming (Chen, Zhou et al. 2024;Mu, Du et al. 2021), game theory(Chen and Yang 2024;Jing,

Hui et al. 2019;Wang, Zhao et al. 2022), hierarchical control(Chen, van Arem et al. 2021;Jing, Hui et al. 2022;Liu, Zhuang et al. 2023;Tang, Zhu et al. 2022), virtual queue and platoon control (Gao, Wang et al. 2022;Huang, Zhuang et al. 2019;Zhu, Wang et al. 2024), centralized control(Luo, Li et al. 2023;Yang, Zhan et al. 2023), and distributed control (Chen, Wang et al. 2021;Xue, Zhang et al. 2023). Chen, Zhou et al. (2024)proposed a mixed-integer nonlinear programming model to optimize trajectories and merging sequences for multiple vehicles. Jing, Hui et al. (2019) developed an optimization framework and algorithm based on cooperative game theory, deriving optimal trajectories as control strategies by minimizing fuel consumption, passenger comfort, and travel time. Wang, Zhao et al. (2022) integrated priority attributes of special CAV models, types, and lanes into a game cost function to obtain optimal merging sequences, solving longitudinal optimal trajectories using Pontryagin's Maximum Principle. Tang, Zhu et al. (2022)proposed a hierarchical System Optimal Cooperative Merging Control model with Flexible Merging Positions (CMC-FMP) for safe and efficient merging, though it only optimizes minimum total delay without considering comfort or emissions. Zhu, Wang et al. (2024) proposed an upper-level control strategy to coordinate on-ramp traffic by actively creating gaps and forming platoons, combining macro-micro traffic flow models to determine optimal plans.Xue, Zhang et al. (2023) developed a distributed cooperative optimal control algorithm based on platoons, projecting on-ramp platoons onto the mainline to transform two-dimensional cooperation into one-dimensional following control.

Learning-based methods, such as reinforcement learning(Cai, Kong et al. 2024;Chen, Du et al. 2024;Li, Wu et al. 2024;Zhang, Wu et al. 2023;Zhou, Zhuang et al. 2022), show potential for human-like decision-making, typically training vehicles to obtain optimal control strategies. Zhou, Zhuang et al. (2022) proposed a cooperative merging control strategy for CAVs based on Distributed Multi-Agent Deep Deterministic Policy Gradient (MADDPG), optimizing rear-end safety, lateral safety, and energy consumption. Li, Zhou et al. (2023) introduced a deep reinforcement learning-based decision method, incorporating a new Time Difference to Merging (TDTM) indicator combined with Time to Collision (TTC) to evaluate safety and optimize on-ramp merging.

Furthermore, with the development of vehicle-road collaboration and increasing CAV penetration, multi-agent cooperative control has emerged. Research in merging areas focuses on dynamic interactions of CAV clusters and two-way data exchange with infrastructure to address multi-CAV cooperative safety. Hu, Li et al. (2024) proposed GMA-DRL, combining a spatial graph convolutional encoder with multi-head attention and Actor-Critic deep reinforcement learning to capture complex temporal-spatial relationships, though its technical complexity may increase computational costs. Wang, Zhao et al. (2023)proposed a cooperative merging framework based on vehicle-infrastructure collaboration, using centralized induction and distributed autonomous control to improve merging efficiency. Peng, Li et al. (2024)designed a distributed real-time information sharing mechanism based on road segment management units for vehicle-road collaboration, enabling real-time sharing of states and intentions. Peng, Xu et al. (2025) proposed mainline-priority and on-ramp-priority cooperative control strategies, pre-planning trajectories via real-time information sharing, but these focus on single-vehicle-type and single-mainline single-ramp scenarios, ignoring heterogeneous vehicle interactions and multi-mainline complexity.

Research Gaps and Challenges:

Most studies focus on single-mainline single-ramp merging, lacking in-depth exploration of cooperative merging control between multi-mainlines and ramps.

Homogeneous vehicle models dominate, with limited consideration of vehicle heterogeneity (e.g., dynamic differences and energy consumption between large and small vehicles) in highway on-ramp traffic. Preemptive spatiotemporal trajectory ddjustment systems have not fully explored the evolution, quantification, and compensation strategies for spatiotemporal errors during pre-planned trajectory execution.

In summary, optimizing multi-mainline on-ramp merging strategies in connected autonomous driving environments—especially addressing collaborative complexity from mixed vehicle types and enabling efficient spatiotemporal resource allocation in multi-mainline scenarios represents a critical challenge in intelligent transportation. This study aims to partially bridge these gaps by introducing a preemptive collaborative control framework for heterogeneous vehicles, integrating spatiotemporal error analysis and adaptive trajectory replanning.

## 3 Preemptive Collaborative Merging Control Strategy

### 3.1 Safe Spatiotemporal Distance

In traditional transportation systems, due to the lack of effective perception of the driving intentions of preceding and surrounding vehicles, vehicles typically need to maintain a large safe distance from the vehicle ahead to ensure timely braking and collision avoidance in emergency situations. Meanwhile, the game-theoretic decision-making based on local information among vehicles further exacerbates the inefficient utilization and waste of spatiotemporal resources. To achieve rational spatiotemporal resource allocation, this paper aims to optimize vehicle operations through pre-planning and programming of trajectories. However, before executing the programmed trajectories, it is essential to comprehensively evaluate and quantify the spatiotemporal errors in actual driving to ensure the accuracy and feasibility of trajectory programming.

During the execution of pre-programmed trajectories, spatiotemporal deviations often arise from factors such as positioning error, time synchronization error, and trajectory tracking error. This study employs GPS RTK differential positioning technology with a positioning error of 0.02 m; data from the National Time Service Center show that the time synchronization error is less than 3 nanoseconds; simulation analysis based on highway design speed (100 km/h), alignment conditions, and differences in acceleration/deceleration performance of different vehicles indicates a spatiotemporal trajectory tracking error of 0.6 m. Considering that the RSMU (Road Segment Management Unit) programs and uploads trajectories before vehicles enter the segment, the transmission delay (in the order of tens of milliseconds) has negligible impact on the future operation states of vehicles, thus it is not considered. The above error parameters provide a quantitative basis for subsequent trajectory programming. Overall, the calculation formula for the safe spatiotemporal spacing that should be maintained between preceding and following vehicles is:

$$S_{safe} = 2\left(L_{pos} + L_{trk}\right) \quad (1)$$

Where $L_{pos}$ reprents GPS positioning error, taking 0.02 m; $L_{trk}$ reprents spatiotemporal trajectory tracking error, taking 0.6 m.

During the merging process, to ensure that ramp vehicles can merge into the mainline traffic flow in a high - speed and safe manner, sufficient safety gaps need to be reserved between two mainline vehicles. The minimum safety gap available for ramp vehicles to merge should meet the following condition: the length of the gap should be at least the sum of twice the safe spatiotemporal spacing and the body length of the ramp vehicle, so as to fully ensure the safety and smoothness of the merging operation. Its calculation method is as follows:

$$L_{safe} = 2S_{safe} + L_v \quad . \tag{2}$$

Where $S_{safe}$ reprents safe spatiotemporal distance maintained between the preceding and following vehicles, in meters; $L_v$ reprents body length of the target vehicle on the ramp. When the vehicle type is CAV, it is set to 5 m; when the vehicle type is CAT, it is set to 7 m.

**3.2 Cooperative Control Strategy**

3.2.1 Cooperative Control Judgment Flow

In the complex traffic scenario where the mainline intersects with the on-ramp, before vehicles enter the merging area and come under the jurisdiction of the Road Segment Management Unit (RSMU), monitoring devices will collect relevant information about road infrastructure in real time and transmit it to the RSMU. Meanwhile, the Vehicle Intelligent Unit (VIU) uploads multi-source information such as its own status, sensor data, and driving intention. The RSMU conducts comprehensive analysis and dynamic prediction based on the received data, formulates the optimal driving trajectory in advance, and shares the planning results with all vehicles within its jurisdiction in real time. Vehicles within the jurisdiction operate in coordination according to the trajectory provided by the RSMU, thereby achieving efficient and stable control of the traffic flow.

Among them, to minimize the conflict risk and optimize the utilization of spatiotemporal resources, it is necessary to ensure that the time difference between the target vehicle on the on-ramp and the cooperative vehicle in the outer lane of the mainline is as small as possible when they reach the preset merging point, while maintaining a safe spatiotemporal spacing between them. Specifically, this paper follows the following principles when selecting cooperative vehicles: In the outer lane of the mainline, priority is given to selecting a vehicle under the condition of traveling at the original speed, for which the time difference between its arrival time at the preset merging point and the time when the on-ramp vehicle departs from the normal section of the on-ramp, first travels at the original speed to the starting point of the parallel acceleration lane, and then accelerates to the preset merging point at the predetermined acceleration (with the speed being consistent with that of the cooperative vehicle on the mainline) is the smallest.

On this basis, the RSMU formulates specific cooperative strategies and pre-plans the spatiotemporal trajectories of vehicles in combination with the evaluation results of the time difference and safe spacing as well as the vehicle types. Measures such as adjusting vehicle speeds and having cooperative vehicles on the mainline change lanes in advance are included to ensure that vehicles can complete the merging process in a safe, smooth, and efficient manner.

3.2.2 On-ramp Target Connected Automated Truck (CAT)- Mainline Cooperative Connected Automated Vehicle (CAV)

When the target vehicle on the on-ramp is a Connected Automated Truck (CAT) and the cooperative vehicle on the mainline is a Connected Automated Vehicle (CAV), considering the differences between CAT and CAV in physical characteristics, maneuvering performance, and safe driving requirements, CAT typically requires a larger acceptable merging gap during the merging process. To address this, the following cooperative merging control strategy is implemented:

If lane-changing conditions are satisfied, prioritize prompting the CAV to change lanes, and coordinate with its preceding and following vehicles to reserve the minimum acceptable safe merging gap for the CAT, proactively optimizing merging conditions and reducing risks. If lane - changing conditions are not met, dynamically adjust the distance between the CAV and its preceding/following vehicles with precision based on the sequence of the two vehicles arriving at the merging point, ensuring the safe spacing between the CAT and mainline vehicles during merging and maintaining traffic flow stability.

Additionally, when the mainline traffic flow is heavy, flexibly adjust the acceleration of the on-ramp CAT to reduce interference with the mainline, as shown in Fig. 1. This approach enhances the operational efficiency and safety of the system in complex merging scenarios.

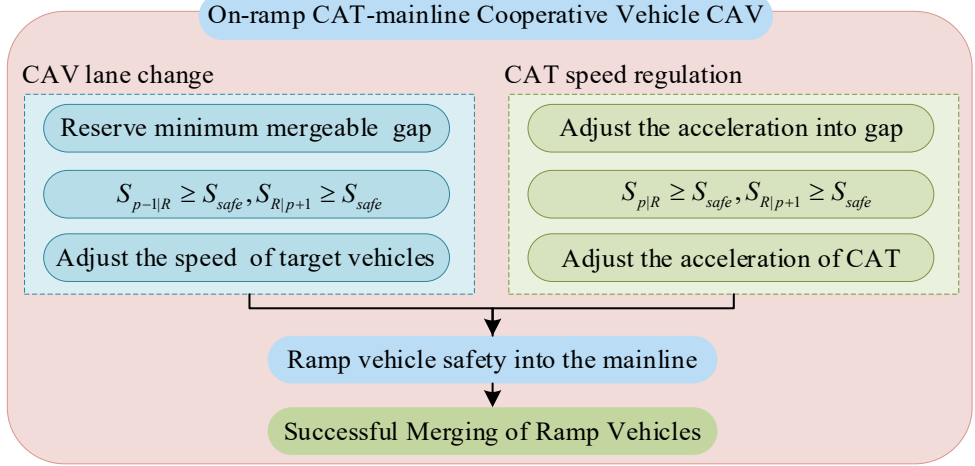

Fig. 1 Merging Strategy for On-ramp Target CAT - Mainline Cooperative CAV

The following elaborates in detail the cooperative control merging concepts under two strategies: prompting mainline CAVs to change lanes and adjusting the speed of on-ramp CATs.

(1) Strategy of Prompting Mainline CAVs to Change Lanes

When the system detects that the on-ramp target vehicle is a CAT and locates the cooperative vehicle in the mainline outer lane as a CAV, it prioritizes prompting the mainline cooperative CAV to implement and complete lane-changing before the start of the parallel acceleration lane. This action aims to enable the preceding and following vehicles (q, q+1, regardless of type) of the mainline cooperative CAV to reserve the minimum safe merging gap $L_{safe}^{CAT}$ required for the CAT, thereby allowing the CAT to safely merge into the mainline traffic flow within the cooperative merging zone.

When the cooperative vehicle p in the outer lane changes lanes to the inner lane, it must consider the spacing with the preceding and following vehicles (q, q+1) in the inner lane $S_f$ and $S_b$, ensuring that the safe distance and speed conditions for lane-changing are met. The positional relationship between the mainline outer-lane cooperative vehicle p and the preceding/following vehicles (q, q+1) in the inner lane before lane-changing is shown in Fig. 2.

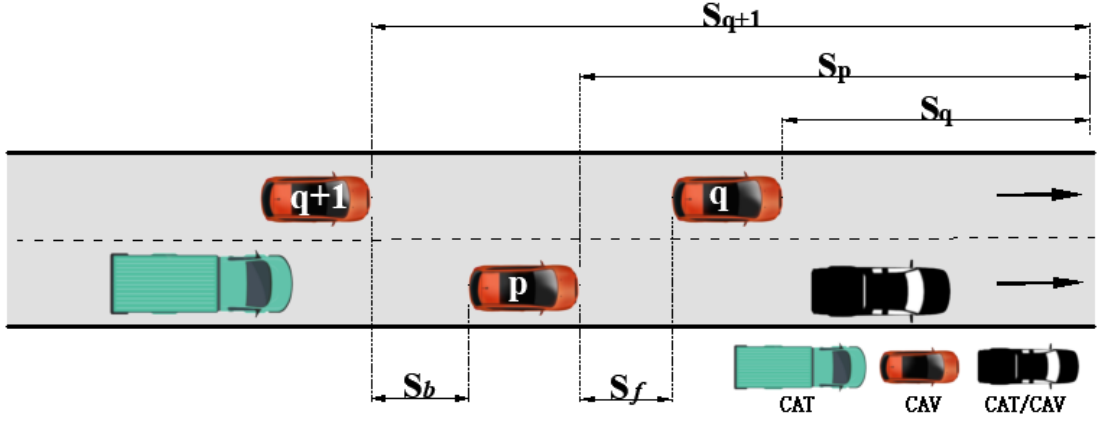

**Fig. 2 Schematic Diagram of Pos Relationship Before Lane – changing**

1) When $S_b \geq S_{safe}$、$S_f \geq S_{safe}$, the cooperative vehicle $p$ directly changes lanes to the inner lane.

2) When $S_b < S_{safe}$, $S_f \geq S_{safe}$, the cooperative vehicle $p$ and the preceding vehicle $q$ in the inner lane travel at a constant speed at the original speed, while the following vehicle $q+1$ in the inner lane travels at a reduced speed. Since the travel speed of the inner lane is higher than that of the outer lane, to minimize the impact of speed fluctuations and reduced traffic efficiency in the inner lane caused by the lane - changing of the cooperative vehicle $p$, it is necessary to determine an appropriate speed adjustment amount. This ensures that when the cooperative vehicle p changes lanes, it can meet the safety gap requirements, and at the same time, be as close as possible to the speed constraint conditions. Moreover, the entire lane - changing process needs to be completed before the start point of the parallel acceleration lane (at the position of 0). The lane - changing process is shown in Fig. 3.

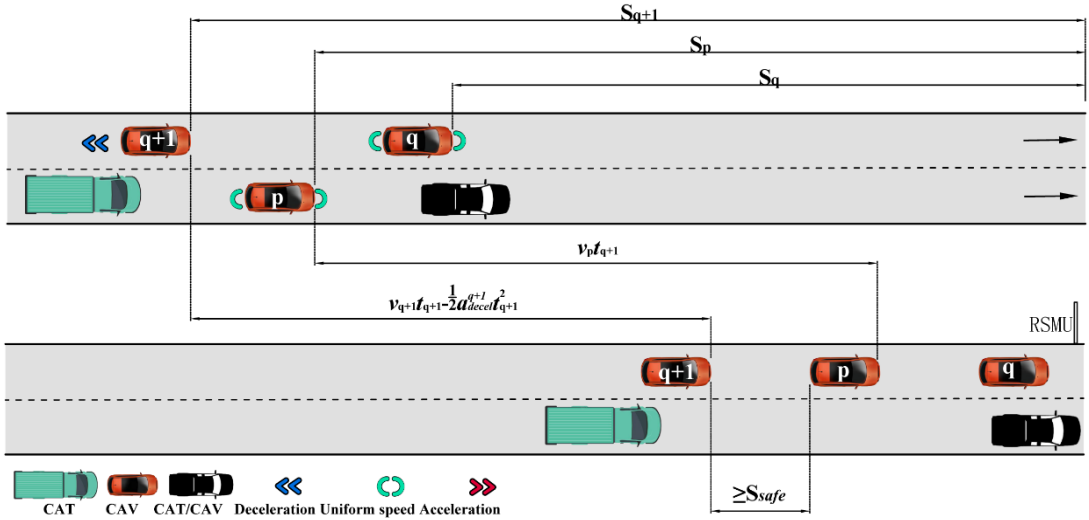

**Fig. 3 Schematic Diagram of Rear Vehicle $q+1$ in the Inner Lane of the Mainline Decelerating to Assist Vehicle p in Lane – changing**

After the lane - changing is completed, the positional relationship shall satisfy the conditions $S'_{q+1} - S'_p - L_{CAV} \geq S_{safe}$ and $S'_p \geq 0$, that is, it shall satisfy:

$$\begin{cases} \left(S_{q+1} - \left(v_{q+1}t_{q+1} - \frac{1}{2}a_{decel}^{q+1}t_{q+1}^2\right)\right) - \left(S_p - v_p t_{q+1}\right) - L_{CAV} \geq S_{safe} \\ 0 < t_{q+1} \leq \dfrac{S_p}{v_p} \end{cases} \quad (3)$$

Where $v_p$、$v_q$ and $v_{q+1}$ are respectively the speeds of the cooperative vehicle p, the preceding vehicle q in the inner lane, and the rear vehicle $q+1$ in the inner lane immediately before strategy is to be implemented, m/s. $a_{decel}^{q+1}$ reprents the deceleration of the rear vehicle q + 1 in the inner lane m/s²; $t_{q+1}$ reprents the deceleration time of the rear car $q+1$ in the inner lane, that is, the lane change cooperation time, s; $S_{safe}$ reprents the safe distance between two vehicles, m; $L_{CAV}$ reprents CAV body length, 5m; $S_p$、$S_{q+1}$ respectively reprent the distance from the cooperative vehicle $p$ and the rear vehicle $q+1$ of the inner lane to the starting point of the acceleration lane when the lane change strategy is about to be implemented, respectively, m ; $S'_p$、$S'_{q+1}$ respectively reprent the distance from the cooperative vehicle $p$ after lane change and the distance from the rear vehicle $q+1$ of the inner lane to the starting point of the acceleration lane, respectively, m.

In summary, the requirements of the rear car $q+1$ in the inner lane according to the deceleration are as follows :

$$a_{decel}^{q+1} \geq \frac{2\left(v_{q+1}t_{q+1} - v_p t_{q+1} - S_{q+1} + S_p + L_{CAV} + S_{safe}\right)}{t_{q+1}^2}. \quad (4)$$

Under the above conditions, take $\min\{a_{decel}^{q+1}\}$, let the rear vehicle $q+1$ of the inner lane slow down according to the deceleration, and reserve a safe distance for the cooperative vehicle.

4) When $S_b \geq S_{safe}$ and $S_f < S_{safe}$, the cooperative vehicle p and the rear vehicle $q+1$ on the inner lane travel at a constant speed according to the original speed, and the front vehicle q on the inner lane accelerates. Because the speed of the rear vehicle $q+1$ in the inner lane is greater than the speed of the cooperative vehicle p in the outer lane, in order to avoid the rear vehicle $q+1$ in the inner lane catches up with the cooperative vehicle p in the time range of the front vehicle q acceleration in the inner lane, which would lead to $S_b < S_{safe}$ and the whole lane change process needs to be completed before the starting point of the parallel acceleration lane ( position is 0 ). The lane change process is shown in Fig. 4:

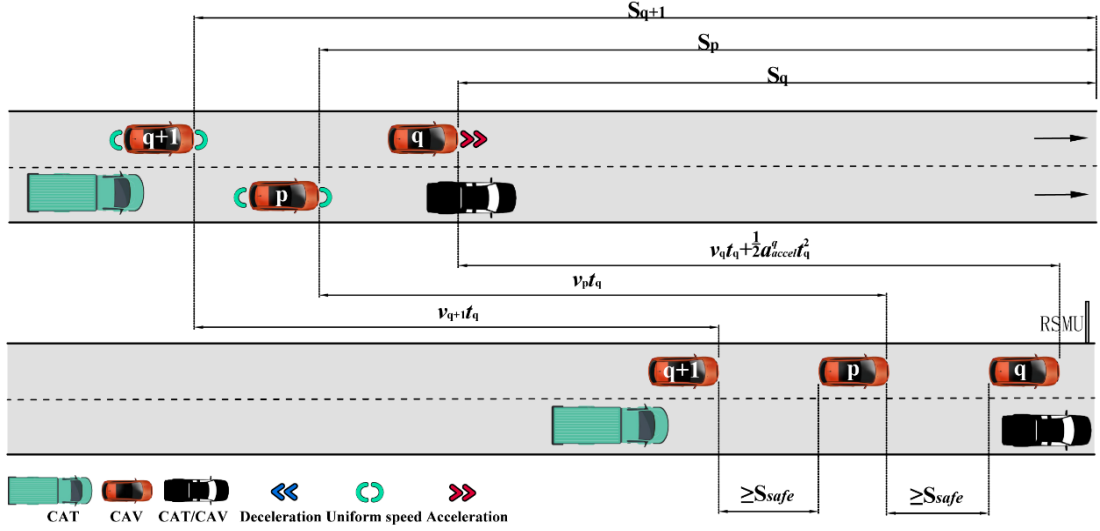

**Fig. 4 The acceleration of the front vehicle q in the inner lane of the main line to assist p to change lanes**

After the lane change is completed, the positional relationship must satisfy $S'_{q+1} - S'_p - L_{CAV} \geq S_{safe}$、 $S'_p - S'_q - L_{CAV} \geq S_{safe}$ and $S'_p \geq 0$. That is, it must satisfy:

$$\begin{cases} S_{q+1} - v_{q+1}t_q - (S_p - v_p t_q) - L_{CAV} \geq S_{safe} \\ S_p - v_p t_q - \left(S_q - \left(v_q t_q + \frac{1}{2}a^q_{accel}t_q^2\right)\right) - L_{CAV} \geq S_{safe} \\ 0 < t_q \leq \dfrac{S_p}{v_p} \\ a^q_{min} \leq a^q_{accel} \leq a^q_{safe} \end{cases} \quad (5)$$

Where $a^q_{accel}$ reprents acceleration of the front vehicle $q$ in the inner lane, m/s²; $a^q_{safe}$ reprents the maximum safe following acceleration of the front vehicle $q$ following the front vehicle in the inner lane, m/s²; $t_q$ reprents the acceleration time of the front vehicle $q$ in the inner lane, that is, the lane change cooperation time, s;

In summary, the $q$ acceleration time $t_q$ of the front vehicle in the inner lane needs to meet $t_q \leq \dfrac{S_{q+1} - S_p - S_{safe} - L_{CAV}}{v_{q+1} - v_p}$ and $0 \leq t_q \leq \dfrac{S_p}{v_p}$ conditions at the same time. The conditions that the $q$ acceleration of the front car $a^q_{accel}$ in the inner lane needs to meet are as follows:

$$a^q_{accel} \geq \frac{2(S_q + S_{safe} + L_{CAV} - S_p + (v_p - v_q)t_q)}{t_q^2}. \quad (6)$$

Under the above conditions, take $\min\{a^q_{accel}\}$, let the front vehicle q of the inner lane accelerate according to the acceleration, and reserve a safe distance for the cooperative vehicle *p*.

In view of the above three situations, the cooperative vehicle p can smoothly perform the lane change operation, and maintain its original speed for uniform driving during the whole lane change process. Although the cooperative vehicle $p$ (CAV) has set aside the necessary space distance for the target vehicle ( CAT ) in advance during the lane change process, in view of the fact that the target vehicle R (CAT) merging into the ramp has a larger body size than the cooperative vehicle $p$ (CAV), it is still necessary to carefully evaluate whether the front and rear vehicles $p$-1 and $p$+1 can provide sufficient minimum merging safety gap for CAT at the merging moment after the successful lane change of the cooperative vehicle p, so as to ensure the smoothness and safety of the merging process. Therefore, the position relationship between the vehicle $p$-1 and $p$+1 and the ramp target vehicle R at the confluence time needs to meet the requirements shown in Fig. 5:

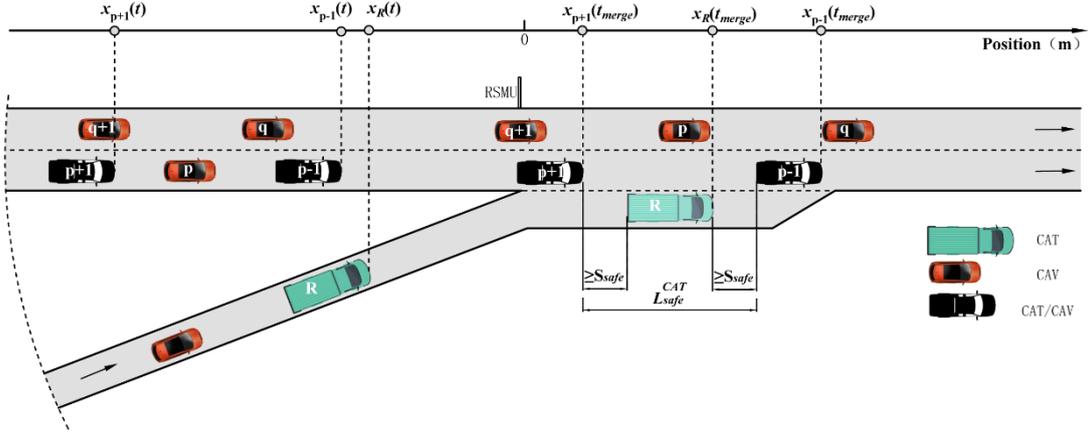

**Fig. 5 The position relationship between the vehicle p-1, p + 1 and the ramp target vehicle R at the confluence time.**

If the system detects that there is no leading vehicle p-1 before the cooperative vehicle p and there is a following vehicle $p$+1 after the cooperative vehicle p, it is only necessary to make the distance between the target vehicle R and the following vehicle $p$+1 on the ramp at the confluence time meet the requirement of greater than or equal to the safe distance $S_{safe}$. If the system monitors that the cooperative vehicle p has both the leading vehicle p-1 and the following vehicle $p$+1, the position relationship of the vehicle $p$-1, $p$+1 and the ramp target vehicle R at the confluence time will appear in the following three situations :

Case 1 : If $S_{p-1|R} < S_{safe}$ and $S_{R|p+1} \geq S_{safe}$, let the leading vehicle $p$-1 accelerate forward $\triangle S_{p-1|R}$.

$$\triangle S_{p-1|R} = S_{safe} - S_{p-1|R}, \tag{7}$$

$$S_{p-1|R} = x_{p-1}(t_{merge}) - x_R(t_{merge}) - L_v, \tag{8}$$

$$a_{accel}^{p-1} = \frac{\triangle S_{p-1|R}}{t_r}. \tag{9}$$

$\triangle S_{p-1|R}$ reprents the leading vehicle $p$-1 needs more driving distance，m; $S_{p-1|R}$ reprents the distance between the leading vehicle $p$-1 ( running at the original speed ) and the ramp target vehicle R at the confluence time, m ; $x_{p-1}(t_{merge})$ reprents the position of the leading vehicle $p$-1 ( original speed ) at the confluence moment, m; $x_R(t_{merge})$ reprents the location of the target vehicle R on the ramp at the merging time, m; $t_r$ reprents the total time of the target vehicle R driving to the preset confluence point, s; $L_v$ reprents the length of the body is determined by the actual type of the leading vehicle *p*-1. When the type is CAV, take 5m, and when the type is CAT, take 7m; $a_{accel}^{p-1}$ reprents acceleration required for leading vehicle *p*-1 driving, m/s².

Case 2 : If $S_{p-1|R} \geq S_{safe}$ and $S_{R|p+1} < S_{safe}$ Let the following vehicle *p*+1 decelerate backwards $\triangle S_{R|p+1}$.

$$\triangle S_{R|p+1} = S_{safe} - S_{R|p+1}, \tag{10}$$

$$S_{R|p+1} = x_R(t_{merge}) - x_{p+1}(t_{merge}) - L_{CAT}, \tag{11}$$

$$a_{decel}^{p+1} = \frac{\triangle S_{R|p+1}}{t_r}. \tag{12}$$

Where $\triangle S_{R|p+1}$ reprents follow the vehicle *p*+1 need less distance, m; $S_{R|p+1}$ reprents the distance between the following vehicle *p*+1 ( running at the original speed ) and the ramp target vehicle R at the confluence time, m ; $x_{p+1}(t_{merge})$ reprents confluence moment follows the position of vehicle *p*+1 ( original speed driving ), m ; $a_{decel}^{p+1}$ reprents -the acceleration required to follow the vehicle *p*+1 driving $S_{R|p+1}$,m/s²;

Case 3 :If $S_{p-1|R} < S_{safe}$ and $S_{R|p+1} < S_{safe}$, let the leading vehicle *p*-1 accelerate forward, and let the following vehicle *p*+1 decelerate backward.

4 ) When $S_b < S_{safe}$ and $S_f < S_{safe}$ , in order to effectively avoid the large fluctuation of vehicle speed in the fast lane ( i.e., the inner lane of the main line ) caused by the demand for lane change operation, and the resulting decline in the overall traffic efficiency of the inner lane, it is considered to directly adjust the cooperative vehicle *p* on the outer lane of the mainline. By accelerating or decelerating, the minimum safe entry clearance $L_{safe}^{CAT}$ required for the target vehicle ( CAT ) on the ramp is reserved in advance.

It is assumed that the time when the target vehicle R on the ramp reaches the preset confluence point is $t_r$, and the time when the cooperative vehicle *p* on the outer lane of the main line reaches the preset confluence point at the original speed is $t_m$. When $t_r > t_m$, it is shown that the cooperative vehicle *p* in the outer lane of the main line reaches the preset confluence point before the ramp target vehicle R, but the difference Δ*t* between the two may be so small that the safety distance $S_{safe}$ at the confluence does not meet the requirements. Therefore, the cooperative vehicle p in the outer lane of the main line is allowed to accelerate and set aside the safe distance required by the ramp target vehicle R. Finally, in the confluence, the distance between the target vehicle R on the ramp and the cooperative vehicle *p* on the outer lane of the main line and its rear vehicle *p*+1 needs to meet the conditions shown in Fig. 6:

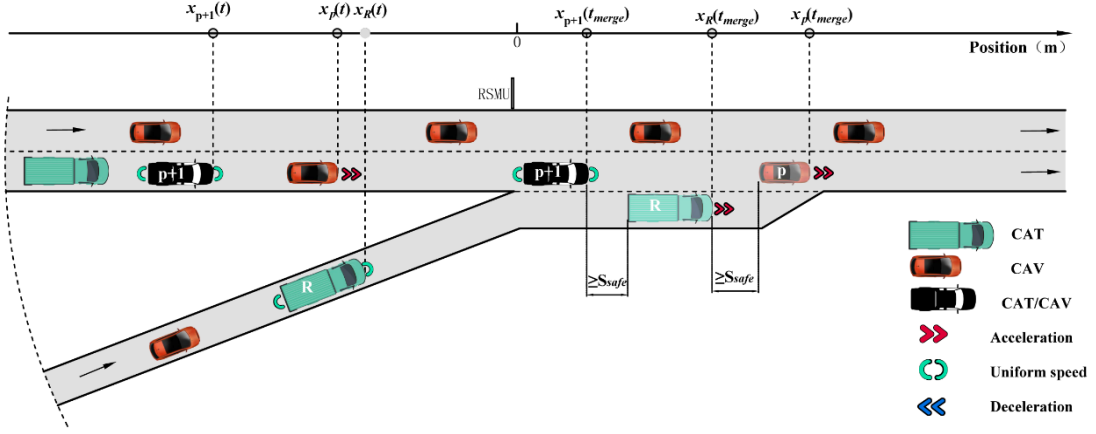

**Fig. 6 Confluence process ( p acceleration )**

The CAT travels from the initial position $x_R(t)$ on the on - ramp at the original speed $v_0^{CAT}$ to the start - point of the acceleration lane, with the travel time being $t_{r_{normal}}^{CAT}$ ; then, it accelerates at $a_{r_{accel}}^{CAT}$ on the acceleration lane to the merging speed $v_{merge}$, and the acceleration time is $t_{r_{accel}}^{CAT}$. The motion process of the on - ramp target vehicle R (CAT) is as follows:

$$v_{merge} = v_0^{CAT} + a_{r_{accel}}^{CAT} t_{r_{accel}}^{CAT}, \tag{13}$$

$$x_R(t_{merge}) = v_0^{CAT} t_{r_{accel}}^{CAT} + \frac{1}{2} a_{r_{accel}}^{CAT} \left(t_{r_{accel}}^{CAT}\right)^2. \tag{14}$$

The following conditions must be satisfied :

$$\begin{cases} x_R(t_{merge}) \leq L_b \\ v_{\lim_{min}} \leq v_{merge} \leq v_{\lim_{max}} \\ a_{\min}^{CAT} \leq a_{r_{accel}}^{CAT} \leq a_{\max}^{CAT} \end{cases}. \tag{15}$$

That is CAT acceleration time on the acceleration lane :

$$\frac{v_{\lim_{min}} - v_0^{CAT}}{a_{r_{accel}}^{CAT}} \leq t_{r_{accel}}^{CAT} \leq \frac{\sqrt{\left(v_0^{CAT}\right)^2 + 2a_{r_{accel}}^{CAT} L_b} - v_0^{CAT}}{a_{r_{accel}}^{CAT}}. \tag{16}$$

$v_{merge}$ reprents the speed of the mainline and ramp vehicles at the time of confluence,m/s; $x_R(t_{merge})$ reprents the location of the target vehicle on the ramp at the merging time,m; $v_0^{CAT}$ reprents CAT speed at the starting point of the acceleration lane,m/s; $a_{r_{accel}}^{CAT}$ reprents CAT acceleration on the acceleration lane,m/s²; $v_{\lim_{min}}$ reprents the minimum speed limit of the outer lane of the main line is

$\left(\dfrac{60}{3.6}\right)$ m/s; $L_b$ reprents the length of parallel acceleration lane is 200 m; $t_{r_{accel}}^{CAT}$ reprents CAT acceleration time on the acceleration lane, s.

The $p$ (CAV) acceleration process of the cooperative vehicle in the outer lane of the main line is:

$$v_{merge} = v_p + a_{accel}^p t_r, \tag{17}$$

$$x_p(t_{merge}) = x_p(t) + v_p t_r + \frac{1}{2} a_{accel}^p t_r^2. \tag{18}$$

Among them:

$$\begin{cases} t_r = t_{r_{normal}}^{CAT} + t_{r_{accel}}^{CAT} \\ t_{r_{normal}}^{CAT} = \dfrac{|x_R(t)|}{v_0^{CAT}} \end{cases}. \tag{19}$$

Where $t_{r_{normal}}^{CAT}$ reprents the driving time of -CAT from the initial position to the starting point of the acceleration lane in the normal section of the ramp CAT, s; $x_R(t)$ reprents the initial position of the ramp target vehicle, m.

The $p+1$ (any type of vehicle) motion process of the rear vehicle of the cooperative vehicle in the outer lane of the main line:

$$x_{p+1}(t_{merge}) = x_{p+1}(t) + v_{p+1} t_r. \tag{20}$$

The constraints that the whole process needs to meet are:

$$\begin{cases} x_p(t_{merge}) - x_R(t_{merge}) - L_{CAV} \geq S_{safe} \\ x_R(t_{merge}) - x_{p+1}(t_{merge}) - L_{CAT} \geq S_{safe} \end{cases}. \tag{21}$$

By substituting formula (4.14), formula (4.18), formula (4.19) and formula (4.20) into (4.21), the acceleration time of CAT on the acceleration lane needs to meet the following conditions:

$$t_{r_{accel}}^{CAT} \geq \frac{\sqrt{(v_0^{CAT} - v_{p+1})^2 + 2a_{r_{accel}}^{CAT}\left(x_{p+1}(t) + L_{CAT} + S_{safe} + v_{p+1} \dfrac{|x_R(t)|}{v_0^{CAT}}\right)} - (v_0^{CAT} - v_{p+1})}{a_{r_{accel}}^{CAT}}. \tag{22}$$

At the same time, the acceleration of the cooperative vehicle p in the outer lane of the main line needs to meet the following conditions:

$$a_{accel}^p \geq \frac{2\left(v_0^{CAT} t_{r_{accel}}^{CAT} + \dfrac{1}{2} a_{r_{accel}}^{CAT}(t_{r_{accel}}^{CAT})^2 + S_{safe} + L_{CAV} - x_p(t) - v_p\left(\dfrac{|x_R(t)|}{v_0^{CAT}} + t_{r_{accel}}^{CAT}\right)\right)}{\left(\dfrac{|x_R(t)|}{v_0^{CAT}} + t_{r_{accel}}^{CAT}\right)^2}. \tag{23}$$

In the satisfiability constraint Eq.(23), $\min\{a_{accel}^p\}$ is taken as the acceleration of the cooperative vehicle p in the outer lane of the main line.

（2）Considering the adjustment of ramp CAT acceleration

When the type of target vehicle entering the system ramp is CAT, the cooperative vehicle in the outer lane of the main line is CAV, and the main line traffic flow is large, in order to reduce the interference to the main line traffic flow, the ramp CAT acceleration is considered to be adjusted. In this case, because the speed of the cooperative vehicle p in the outer lane of the main line is not adjusted, it runs at a constant speed to the confluence point, that is, $v_{merge} = v_p$. The confluence process is shown in Fig. 7:

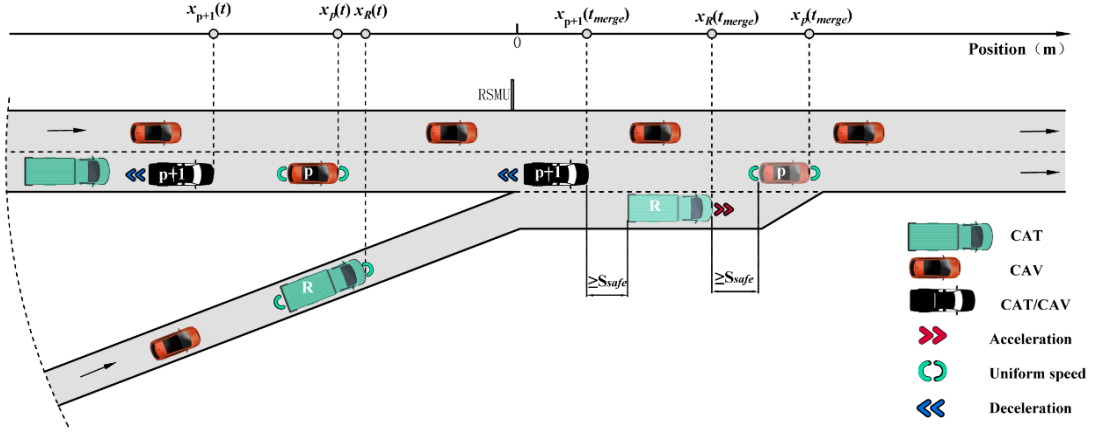

Fig. 7 Adjustment ramp R acceleration confluence process diagram

The acceleration process of the ramp target vehicle R is

$$x_R(t_{merge}) = \frac{v_p^2 - (v_0^{CAT})^2}{2a'_{r_{accel}}}. \qquad (24)$$

The motion process of the cooperative vehicle p in the outer lane of the main line is：

$$x_p(t_{merge}) = x_p(t) + v_p \left( \frac{|x_R(t)|}{v_0^{CAT}} + \frac{v_p - v_0^{CAT}}{a'_{r_{accel}}} \right). \qquad (25)$$

The conditions that the whole process needs to meet：

$$x_p(t_{merge}) - x_R(t_{merge}) - L_{CAV} \geq S_{safe}. \qquad (26)$$

Substituting Eq.(24) and Eq.(25) into Eq.(26), the condition that the ramp adjustment acceleration $a'_{r_{accel}}$ satisfies is

$$a'_{r_{accel}} \leq \frac{(v_p - v_0^{CAT})^2}{2\left( S_{safe} + L_{CAV} - x_p(t) - v_p \frac{|x_R(t)|}{v_0^{CAT}} \right)}. \qquad (27)$$

At the same time, it is also necessary to ensure that the distance between the rear vehicle *p+1* of the cooperative vehicle on the outer lane of the main line and the target vehicle R on the ramp meets the requirements. If the following car *p+1* travels at the original speed to the confluence point $S_{R|p+1} < S_{safe}$, the following car *p+1* is considered to slow down. The motion process and the conditions to be satisfied are as follows :

$$x_{p+1}\left(t_{merge}\right) = x_{p+1}(t) + v_{p+1}\left(\frac{|x_R(t)|}{v_0^{CAT}} + \frac{v_p - v_0^{CAT}}{a'_{r_{accel}}}\right) - \frac{1}{2}a_{decel}^{p+1}\left(\frac{|x_R(t)|}{v_0^{CAT}} + \frac{v_p - v_0^{CAT}}{a'_{r_{accel}}}\right)^2. \quad (28)$$

$$x_R\left(t_{merge}\right) - x_{p+1}\left(t_{merge}\right) - L_{CAT} \geq S_{safe}. \quad (29)$$

By substituting formula (25) and formula (28) into formula (29), the rear car *p*+1 deceleration $a_{decel}^{p+1}$ can be obtained:

$$a_{decel}^{p+1} \geq \frac{2\left(S_{safe} + L_{CAT} + x_{p+1}(t) + v_{p+1}\left(\frac{|x_R(t)|}{v_0^{CAT}} + \frac{v_p - v_0^{CAT}}{a'_{r_{accel}}}\right) - \frac{v_p^2 - \left(v_0^{CAT}\right)^2}{2a'_{r_{accel}}}\right)}{\left(\frac{|x_R(t)|}{v_0^{CAT}} + \frac{v_p - v_0^{CAT}}{a'_{r_{accel}}}\right)^2}. \quad (30)$$

### 3.3 Ramp Target CAT - Mainline Collaboration CAT

In view of the fact that this paper assumes that the inner lane of the main line only allows CAV to pass based on different vehicle lane changing characteristics, the lane changing strategy is no longer adopted when the cooperative vehicle for the positioning of the outer lane of the main line is CAT, but the speed regulation strategy is the main adjustment method.

When the type of target vehicle entering the system ramp is CAT and the cooperative vehicle in the outer lane of the main line is CAT, in order to reduce the interference to the main line traffic flow, the ramp CAT acceleration is considered to be adjusted. In this case, because the speed of the cooperative vehicle p in the outer lane of the main line is not adjusted, it runs at a constant speed to the confluence point, that is, $v_{merge} = v_p$. The confluence process is shown in Fig. 8:

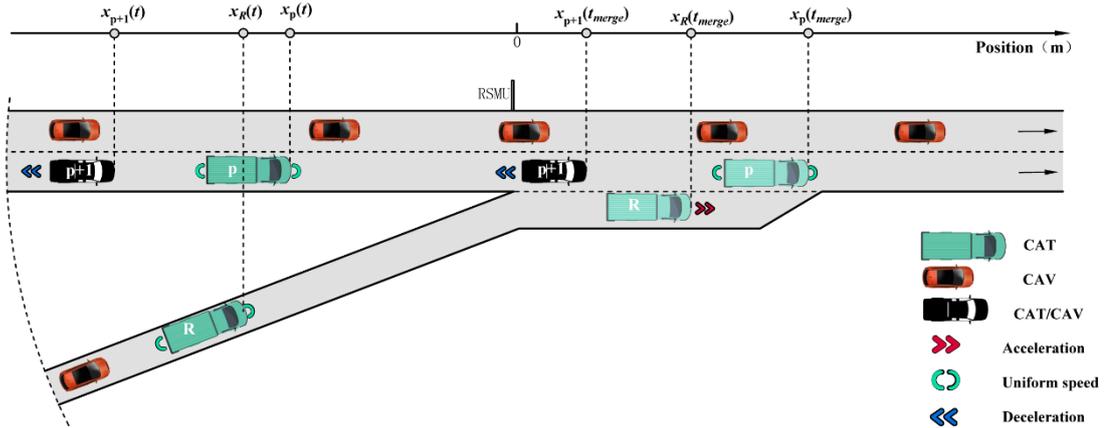

Fig. 8 Adjusting ramp R ( CAT ) acceleration confluence process

The acceleration process of the ramp target vehicle R is listed in Eq. ( 4.30 ), and the motion process of the cooperative vehicle *p* in the outer lane of the main line is listed in Equation ( 4.31 ). However, since the main line cooperative vehicle type is CAT, the whole process needs to meet the following conditions :

$$x_p(t_{merge}) - x_R(t_{merge}) - L_{CAT} \geq S_{safe}. \tag{31}$$

The ramp adjustment acceleration $a'_{r_{accel}}$ satisfies the following conditions:

$$a'_{r_{accel}} \leq \frac{(v_p - v_0^{CAT})^2}{2\left(S_{safe} + L_{CAT} - x_p(t) - v_p \frac{|x_R(t)|}{v_0^{CAT}}\right)}. \tag{32}$$

At the same time, it is also necessary to ensure that the distance between the rear vehicle p+1 of the cooperative vehicle in the outer lane of the main line and the target vehicle R on the ramp meets the requirements. If the time distance from the uniform speed of the rear vehicle to the confluence point does not meet the requirements, the expression of the p+1 deceleration process and the confluence time of the rear vehicle is consistent with the expression and formula Eq.(28) and Eq.(29) of the ramp target vehicle spacing requirements. At this time, the solution of the rear car p+1 deceleration only needs to substitute the result of the formula Eq.(32) into Eq.(30)

## 4 Simulation and evaluation

In order to verify the rationality and effectiveness of the preemptive cooperative confluence control strategy proposed in this paper, which integrates vehicle type, driving intention and space-time safety distance, this chapter constructs the confluence simulation environment of double mainline and single ramp of expressway, and uses the preemptive cooperative confluence control strategy proposed in Chapter 4 to compare the simulation experiment with the uncontrolled situation. The test scheme selects the main line flow of 1000veh / h / lane, 1400veh / h / lane and the ramp flow of 600veh / h / lane for full combination simulation. The vehicle operation characteristics and confluence dynamic differences are qualitatively analyzed by comparing the time-position-speed diagram, and the average speed, average delay and total fuel consumption are used as evaluation indicators to quantitatively evaluate the performance advantages of the preemptive cooperative confluence control strategy.

### 4.1 Simulation scheme setting

In order to systematically evaluate the operation efficiency of the proposed preemptive cooperative confluence control strategy, this paper selects the mainline traffic flow ( 1000,1400veh / h / lane ) and ramp traffic flow (600veh / h / lane ) for full factor combination, and constructs different traffic flow combinations. On this basis, the simulation tests under the preemptive cooperative confluence control condition and the uncontrolled condition are carried out respectively, and the adaptability and operation performance of the proposed strategy under different traffic load conditions are analyzed.

In the case of no control strategy, the longitudinal car-following behavior of the vehicle is simulated by IDM ( Intelligent Driver Model ), and the lateral lane-changing behavior is realized according to the LC2013 lane-changing model. The relevant parameter settings are listed in Table 1. In the preemptive cooperative control strategy, the vehicle type and parameters are consistent with the non-control strategy, but the car-following lane-changing behavior involved in the confluence process is adjusted based on the safe space-time spacing and car-following lane-changing rules proposed in this paper.

**Table 1 No control parameter table**

| Type | Parameter | Meaning | Numerical | Unit |
| --- | --- | --- | --- | --- |
| CAV | length | Vehicle length | 5 | m |
| | accel | Maximum acceleration | 2.5 | m/s² |
| | decel | Maximum deceleration | 4.5 | m/s² |
| CAT | length | Vehicle length | 7 | m |
| | accel | Maximum acceleration | 1.5 | m/s² |
| | decel | Maximum deceleration | 4 | m/s² |
| IDM | mingap | The minimum safety distance between the front and rear vehicles | 2 | m |
| | tau | The minimum time headway expected by the driver | 1 | s |
| LC2013 | lcStrategic | Willingness to change lane | 1 | / |
| | lcCooperative | Lane-changing cooperation | 1 | / |
| | lcSpeedGain | Profit on speed | 1 | / |

### 4.2 Evaluation index selection

Traffic evaluation index can quantitatively characterize the operation characteristics of road traffic flow. Average delay, average speed and total fuel consumption were used for evaluation.

( 1 ) Average delay

The average delay refers to the additional travel time loss caused by the interference of the main line vehicle by other vehicles, which can effectively measure the driving efficiency of the vehicle. The specific calculation of the average delay is :

$$\bar{d} = \frac{1}{n}\sum_{i=1}^{n}\left(t_i^{act} - t_i^{\min}\right), \tag{33}$$

$$t_i^{\min} = \frac{L}{v_{i_{free}}}. \tag{34}$$

Where $\bar{d}$ reprents the average delay per vehicle; $t_i^{act}$ reprents the actual time of the $i$ vehicle leaving the road; $t_i^{\min}$ reprents the minimum time required for the $i$ vehicle to leave the road in the free flow state; $v_{i_{free}}$ reprents the speed of the $i$ vehicle in the free flow state; $L$ reprents the total length of the road section.

(2 ) Average speed

In traffic simulation, the average speed is often regarded as a key indicator to measure the overall traffic flow situation. For a given time interval on a given path ( mainline or ramp ), if n vehicles are observed in that interval and the instantaneous speed $v_i$ ( $i = 1, 2, 3.....n$ ) of each vehicle is recorded, the average speed in that time interval is calculated as follows :

$$\bar{v} = \frac{1}{n}\sum_{i=1}^{n}v_i. \tag{35}$$

Where $v_i$ reprents the instantaneous speed of the $i$ vehicle in this time interval, and $n$ is the total number of vehicles observed in this time interval.

## 4.3 Analysis of simulation results

Aiming at 9 kinds of traffic flow combinations, the trajectory and speed data of vehicles on the outer lane and ramp of the main line are extracted under the control strategy of non-control and preemptive cooperative confluence, and the time-position-speed diagram is drawn to qualitatively analyze the vehicle operation characteristics and confluence dynamic differences under different strategies. The quantitative analysis of the average speed and average delay reveals the optimization effect of the preemptive cooperative confluence control strategy compared with the uncontrolled situation.

4.3.1Time-position-velocity diagram analysis

Under the condition that the main line flow is 1000veh / h / lane, the interference flow of ramp 600veh / h / lane is introduced, and the time-position-speed diagram of the main line lateral lane and ramp vehicle under the non-control and preemptive coordination strategy is extracted. The simulation results are shown in Fig. 9-Fig. 12.

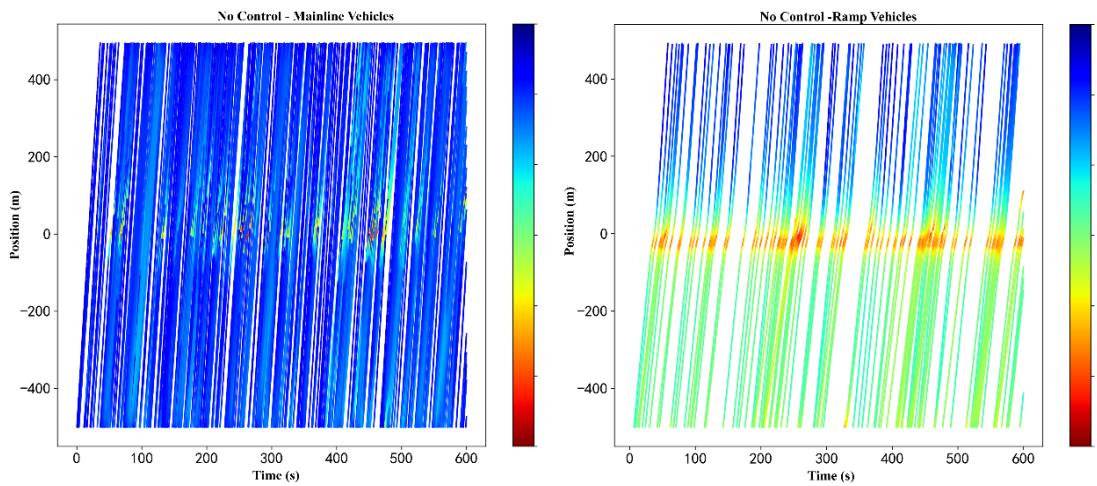

**Fig. 9   Uncontrolled simulation results ( main line : 1000veh / h / lane ; ramp : 600veh / h / lane )**

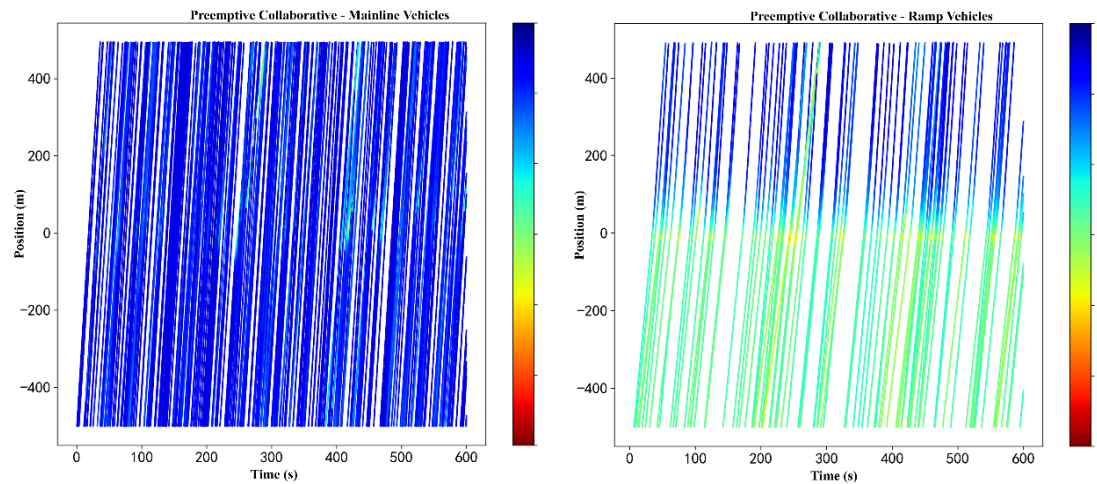

**Fig. 10   Preemptive collaborative simulation results ( main line : 1000veh / h / lane ; ramp : 600veh / h / lane )**

Under the condition that the main line flow is 1400veh / h / lane, the interference flow of ramp 600veh / h / lane is introduced respectively, and the vehicle time-position-speed diagram of the outer lane and ramp of the main line under the non-control and preemptive coordination strategy is extracted. The simulation results are shown in Fig. 11-Fig. 12.

In contrast, after the implementation of the preemptive cooperative confluence control strategy, the overall running state of the vehicle is significantly improved. This strategy effectively improves the driving speed of the mainline and ramp vehicles, and greatly reduces the conflict probability and sudden braking phenomenon between vehicles by planning the merging time in advance and adjusting the vehicle spacing, thus making the traffic flow smoother and more orderly.

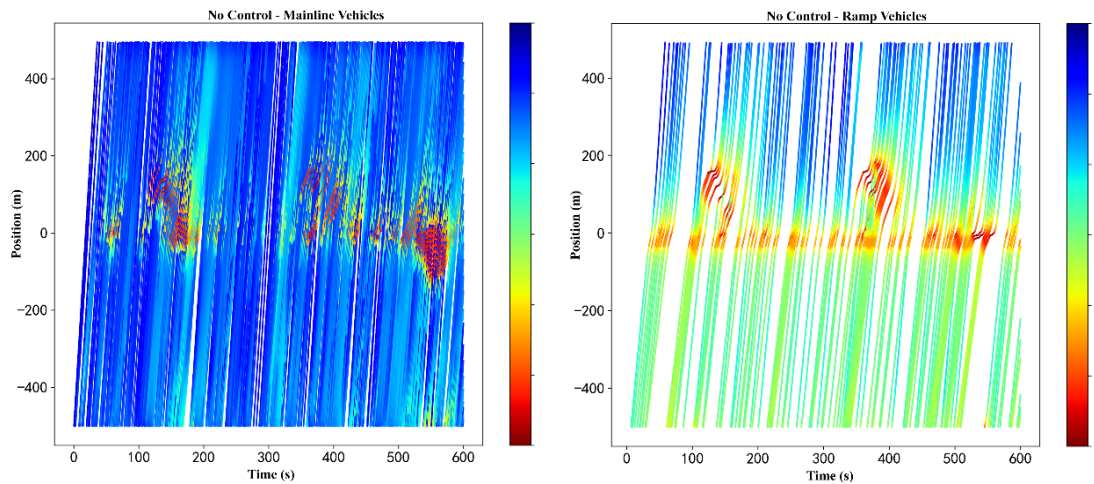

Fig. 11 Uncontrolled simulation results ( main line : 1400veh / h / lane ; ramp : 600veh / h / lane )

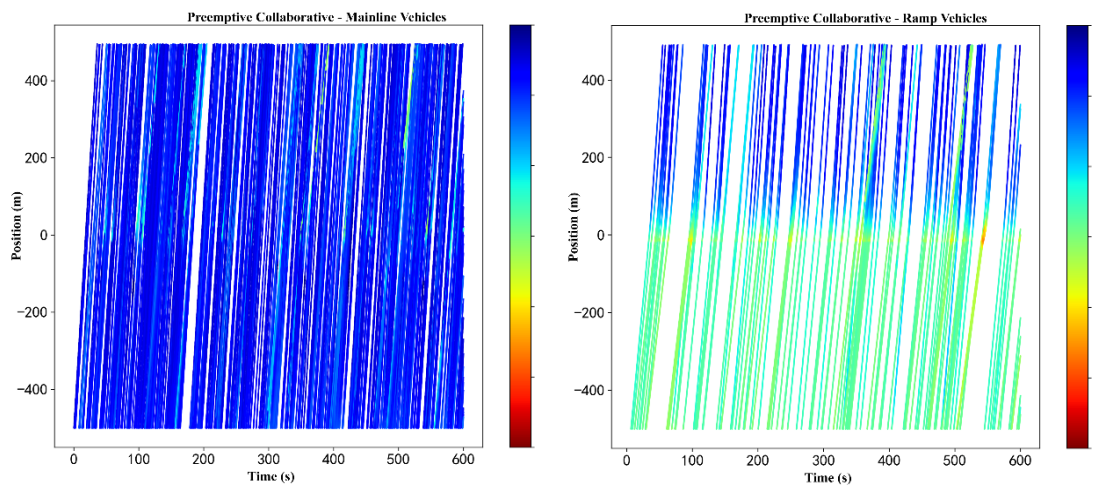

Fig. 12 Preemptive collaborative simulation results ( main line : 1400veh / h / lane ; ramp : 600veh / h / lane )

### 4.3.2 Average speed

This paper takes 0.1s as the simulation time step, collects the average speed data of the main line and ramp vehicles in the simulation cycle, and makes statistical and visual analysis of the data. The relevant results are shown in Fig. 13.

( 1 ) Average speed of mainline vehicle

By selecting the following four groups of working conditions, the average speed of the main line vehicle under the non-control and preemptive cooperative confluence control strategy is compared and analyzed : the main line flow of 1400veh / h / lane and the ramp flow of 600veh / h / lane .

On the whole, after the implementation of the preemptive cooperative control strategy, the average speed of the main line under each working condition is higher than that of the uncontrolled scene, and the speed fluctuation is small, indicating that the strategy can not only improve the traffic efficiency of the main line, but also help to enhance the stability of traffic flow. When the ramp flow is 600veh / h / lane, the mainline flow is 1000veh / h / lane and 1400veh / h / lane, the average speed fluctuation of the mainline vehicle after the implementation of the preemptive coordination strategy is small ; In the uncontrolled scenario, the running speed of the mainline vehicle is significantly reduced, but the preemptive cooperative control strategy can still maintain the optimization advantage of the mainline speed, thus verifying the effectiveness of the strategy in improving the efficiency of the mainline traffic operation. The simulation results are shown in Fig. 13.

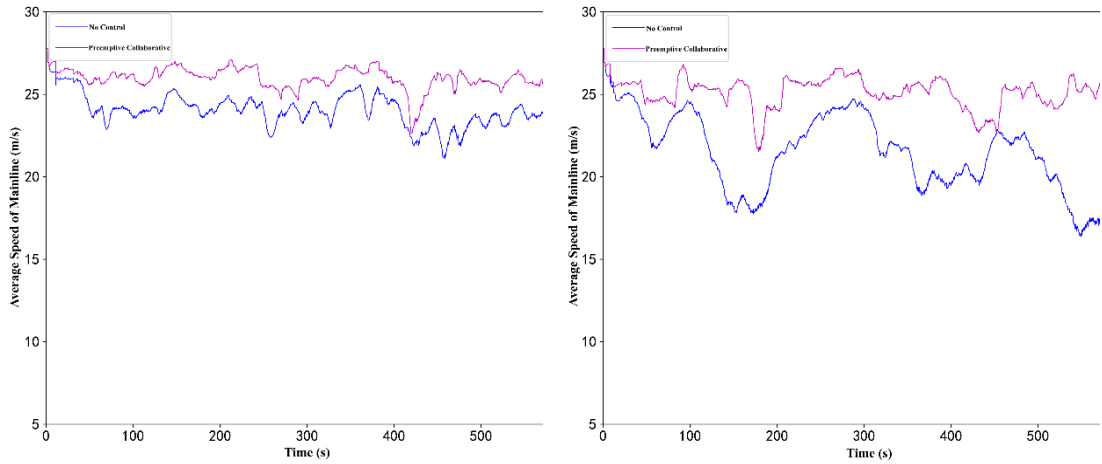

( a ) Mainline 1000-Ramp 600veh / h / lane　　( b ) Mainline 1400-Ramp 600veh / h / lane

**Fig. 13 Comparison of average speed of main line under different traffic flow combinations**

### 4.3.3 Average delay

For the preset nine traffic flow combinations, the average delay time of the mainline and ramp vehicles is collected under the control-free and pre-emptive cooperative confluence control strategies, respectively. The relevant simulation results are listed in Table 5.5 and Table 5.6. The ' improvement rate ' index listed in the table is used to quantitatively measure the improvement effect of preemptive cooperative control on delay optimization compared with no control situation. The calculation formula is as follows :

$$I_{\bar{d}} = \left(1 - \frac{\bar{d}_{pc}}{\bar{d}_{nc}}\right) \times 100\%. \tag{36}$$

Where $I_{\bar{d}}$ reprents the average delay improvement rate; $\bar{d}_{nc}$ reprents the average delay without control; $\bar{d}_{pc}$ reprents the average delay in the case of preemptive coordination.

( 1 ) Average delay of mainline vehicles

The average delay time of mainline vehicles under different traffic flow combinations is listed in Table 2 under the strategy of no control and preemptive coordination.

Table 2　Comparison of average delay time of mainline vehicles under different traffic flow combinations

| Mainline flow /(veh/h/lane) | Ramp flow /(veh/h/lane) | Average delay of main line /(s) | | |
|---|---|---|---|---|
| | | Uncontrolled | Preemptive collaborative | Improvement rate /(%) |
| 1000 | 300 | 3.59 | 0.97 | 72.98 |
| 1000 | 500 | 4.79 | 1.30 | 72.86 |
| 1000 | 600 | 4.39 | 1.22 | 72.21 |
| 1400 | 300 | 6.39 | 1.91 | 70.11 |
| 1400 | 500 | 7.63 | 2.14 | 71.95 |

| Mainline flow /(veh/h/lane) | Ramp flow /(veh/h/lane) | Average delay of main line /(s) | | |
|---|---|---|---|---|
| | | Uncontrolled | Preemptive collaborative | Improvement rate /(%) |
| 1400 | 600 | 9.78 | 1.72 | 82.41 |
| 1800 | 300 | 12.7 | 2.07 | 83.70 |
| 1800 | 500 | 17.63 | 2.33 | 86.78 |
| 1800 | 600 | 22.55 | 2.20 | 90.24 |

From the comparison results of Table 2, it can be seen that the proposed strategy can effectively reduce the average delay of the mainline vehicles in all 9 traffic flow combinations, and the delay improvement rate is up to 90.24 % and the lowest is 70.11 %.
Specifically, when the main line flow is low ( 1000veh / h / lane ) and the ramp flow is 300veh / h / lane, the average delay of the uncontrolled main line is 3.59 seconds, and the first cooperative control is reduced to 0.97 seconds, the improvement rate is 72.98 %, and the traffic flow stability is maintained while reducing the delay. In the case of medium traffic flow ( 1400veh / h / lane ), taking the ramp traffic flow 600veh / h / lane as an example, there is no control delay of 9.78 seconds, and it is reduced to 1.72 seconds after preemptive cooperative control. The improvement rate is increased to 82.41 %, and the optimization effect is further enhanced. When the main line traffic is high ( 1800veh / h / lane ), such as the ramp traffic 600veh / h / lane scenario, the uncontrolled delay is as high as 22.55 seconds, and the preemptive coordinated control is sharply reduced to 2.20 seconds. The improvement rate is 90.24 %, which shows more significant delay reduction effect than the uncontrolled scheme.

The above data show that the proposed preemptive cooperative confluence control strategy has outstanding advantages in high load operation environment, and shows good adaptability with the change of flow rate. In order to more intuitively present the delay of mainline vehicles under different combinations of mainline and ramp flow and the optimization effect after the implementation of preemptive cooperative confluence control strategy, Fig. 14 draws the corresponding histogram of average delay of mainline vehicles.

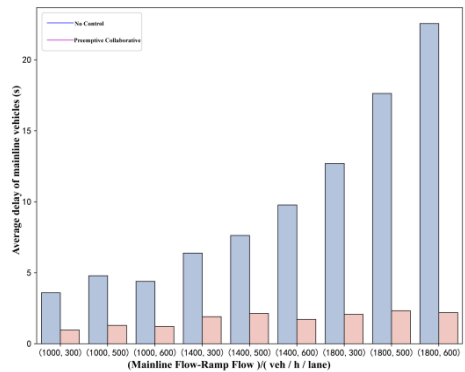

**Fig. 14 Average delay of mainline vehicles**

In summary, on the whole, the delay improvement rate of on-ramp vehicles is generally lower than that of mainline vehicles ( for example, the highest improvement rate of mainline vehicles is 90.24 %, while that of on-ramp vehicles is only 74.24 % ). This phenomenon is mainly due to the fact that the preemptive cooperative confluence control strategy gives priority to ensuring the traffic efficiency of the mainline vehicles when the mainline traffic flow is large, and adjusts the speed of the ramp vehicles to adapt to the rhythm of the mainline traffic flow, thus reflecting the priority of the mainline in the allocation of space-time resources, resulting in relatively limited improvement of ramp delays. Nevertheless, compared with the uncontrolled situation, this strategy can effectively reduce the average delay of the mainline and ramp vehicles under different traffic flow combinations, which verifies its effectiveness in improving the overall traffic efficiency.

## 5 Conclusion

Highway expressway double - main - line and single - ramp confluence scenarios often face challenges in traffic efficiency and safety due to the interaction of different vehicle types and complex traffic flows. To address these issues, this study deduces the calculation method of safe spatio - time distance and proposes a preemptive cooperative confluence control strategy that combines vehicle type, driving intention, and safe spatio - time distance. The main research contents and conclusions are as follows:

From the perspective of coordinating confluence spatio - time resources, a quantitative analysis of the reasonable value of safe spatio - time distance in trajectory pre - preparation is carried out. The safe spatio - time distance is defined as twice the sum of the positioning error and the spatio - time trajectory tracking error. On this basis, it is further clarified that the minimum safety gap required for ramp vehicles to merge into the main line is the sum of twice the safe spatio - time interval and the length of the ramp vehicle body. This provides a theoretical basis for ensuring traffic safety during the confluence process.

A merging simulation platform for heterogeneous vehicles in the dual-mainline single-ramp scenario is built . Through the comparison of time-position-speed diagrams, the vehicle operation characteristics and merging dynamic differences are qualitatively analyzed. Average speed and average delay serve as evaluation indicators for quantitative assessment. Under certain mainline and ramp flow conditions, the proposed strategy can effectively prevent potential vehicle conflicts and emergency braking, enhancing driving safety in the confluence area. It also demonstrates notable benefits in driving stability, overall traffic efficiency, and fuel consumption optimization, with favorable improvement in the average delay of both mainline and ramp vehicles.

This research focuses primarily on the local preemptive cooperative optimization for the merging areas of highway bottleneck on-ramps. In the future, consideration can be given to extending this method to spatiotemporal scheduling optimization on a larger scale. It will cover more complex traffic scenarios and system environments, so as to further enhance the overall efficiency and safety of the highway network.

## CRediT authorship contribution statement

**Yuan Li:** Writing-review & editing, Writing - original draft, Methodology, Investigation, Formal analysis, Supervision.**Xiaoxue Xu:** Writing-review & editing, Methodology, Conceptualization. **Xiang Dong:** Writing - review & editing, Writing-original draft, validation, Software, Investigation. **Junfeng Hao:** Funding acquisition, Validation, Supervision,  Project administration, Investigation. **Tao Li:** Investigation, Formal analysis, Conceptualization. **Sana Ullah:** Visualization, Software. **Chuangrui Huang:** Investigation, Formal analysis, Data curation. **Junjie Niu and Ziyan Zhao:** Formal analysis, Data curation. **Ting Peng:** Funding acquisition, Review, Supervision, Methodology, Conceptualization.